\documentclass{emulateapj}
\slugcomment{9 manuscript pages, including 3 figures. Accepted to
Icarus.}
\shorttitle{Formation of Uranus and Neptune}
\shortauthors{Dodson-Robinson and Bodenheimer}
\begin{document}

\title{The Formation of Uranus and Neptune in Solid-Rich Feeding Zones:
Connecting Chemistry and Dynamics}

\author{Sarah E. Dodson-Robinson}
\affil{University of Texas Astronomy Department \\
1 University Station C1400, Austin, TX 78712}
\email{sdr@astro.as.utexas.edu}

\author{Peter Bodenheimer}
\affil{UCO/Lick Observatory, University of California at Santa Cruz \\
1156 High St, Santa Cruz, CA 95064}

%Proposed running head: Formation of Uranus and Neptune

%\bigskip

%Please direct correspondence to:

%\begin{center}
%\begin{tabular}{l}
%Sarah Dodson-Robinson \\
%University of Texas \\
%Astronomy Department \\
%1 University Station C1400 \\
%Austin, TX 78712 \\
%sdr@astro.as.utexas.edu \\
%\end{tabular}
%\end{center}

%\clearpage

%\begin{center}
%{\bf ABSTRACT}
%\end{center}
\begin{abstract}

The core accretion theory of planet formation has at least two
fundamental problems explaining the origins of Uranus and Neptune: (1)
dynamical times in the trans-Saturnian solar nebula are so long that
core growth can take $> 15$~Myr, and (2) the onset of runaway gas
accretion that begins when cores reach $\sim 10 M_{\oplus}$ necessitates
a sudden gas accretion cutoff just as Uranus and Neptune's cores reach
critical mass. Both problems may be resolved by allowing the ice giants
to migrate outward after their formation in solid-rich feeding zones
with planetesimal surface densities well above the minimum-mass solar
nebula. We present new simulations of the formation of Uranus and
Neptune in the solid-rich disk of \cite{icelines} using the initial
semimajor axis distribution of the Nice model (Gomes et al.\ 2005;
Morbidelli et al.\ 2005; Tsiganis et al.\ 2005), with one ice giant
forming at 12~AU and the other at 15~AU. The innermost ice giant reaches
its present mass after 3.8-4.0~Myr and the outermost after 5.3-6~Myr, a
considerable time decrease from previous one-dimensional simulations
(e.g.\ Pollack et al.\ 1996). The core masses stay subcritical,
eliminating the need for a sudden gas accretion cutoff.

Our calculated carbon mass fractions of $22\%$ are in excellent
agreement with the ice giant interior models of \cite{podolak95} and
\cite{marley95}. Based on the requirement that the ice giant-forming
planetesimals contain $> 10\%$ mass fractions of methane ice, we can
reject any solar system formation model that initially places Uranus and
Neptune inside of Saturn's orbit. We also demonstrate that a large
population of planetesimals must be present in both ice giant feeding
zones throughout the lifetime of the gaseous nebula. This research marks
a substantial step forward in connecting both the dynamical and chemical
aspects of planet formation. Although we cannot say that the solid-rich
solar nebula model of Dodson-Robinson et al.\ (2009) gives {\it exactly}
the appropriate initial conditions for planet formation, rigorous
chemical and dynamical tests have at least revealed it to be a {\it
viable} model of the early solar system.

%Our model's accuracy is limited by our use of the feeding zone
%approximation, in which planetesimals are not strongly scattered. The
%formation timescales we calculate should be viewed as lower limits,
%since planetesimal scattering strongly limits accretion rates in the
%trans-Saturnian solar nebula. In light of the progress we have made on
%reproducing Uranus and Neptune's solid/gas ratios and carbon abundances,
%however, we strongly encourage more detailed N-body investigations of
%embryo growth rates between 12 and 15~AU. 
\end{abstract}

%Icarus keywords
%{\bf Keywords:} planetary formation; Uranus; Neptune; origin, Solar
%System

%\clearpage

\section{Introduction: Statement of the problem and previous work on ice
giant formation}
\label{introduction}

The canonical core accretion theory of planet formation, in which
planetesimals collide to form solid cores which then destabilize the
surrounding gas to accrete an atmosphere (Safronov 1969; Pollack et al.\
1996), has at least two fundamental problems explaining the origins of
Uranus and Neptune.  First, dynamical times in the trans-Saturnian solar
nebula are so long and solid surface densities $\Sigma$ are so low ($<
1$~g~cm$^{-2}$) according to the assumed $\Sigma \propto R^{-2}$ mass
distribution (Pollack et al.\ 1996) that planet growth takes $> 15$~Myr,
far longer than both observed and theoretical protostellar disk
lifetimes (Haisch et al.\ 2001; Alexander et al.\ 2006). Second, runaway
gas accretion begins when solid cores reach 10 to 15~$M_{\oplus}$,
requiring a sudden and complete gas accretion cutoff just as Uranus and
Neptune reach their current masses. \cite{pollack96} pointed out these
problems in their seminal paper on the viability of the core accretion
theory. More recently, Benvenuto et al.\ (2009) showed that Uranus and
Neptune could grow within a few Myr in a population of planetesimals
with a distribution of radii between 30 and 100~km. However,
planetesimals as small as 30~km are not consistent with the prevailing
theory of planetesimal formation, based on the streaming instability,
which produces planetesimals around 100 km and in some cases up to the
radius of Ceres (457 km; Johansen et al.\ 2007).

%However, the fact that the ice giants have $< 10$\% gas mass means some
%type of growth by solid accretion had to occur: gravitational
%instability, even when followed by planetesimal sedimantation and
%envelope photoevaporation, is not a viable formation mechanism (Helled
%et al., in prep).

Uranus and Neptune's total masses, 14.5 and 17.2~$M_{\oplus}$
respectively, place them squarely in the predicted critical mass range
for nucleating an instability in the surrounding gas and accreting
Jupiter's mass or more in under 1000 years (Mizuno 1980; Papaloizou and
Nelson 2005). {\it The first challenge for theorists is to find a
combination of the parameters that control core accretion---feeding zone
location, ice inventory and planetesimal surface density---that leads to
solid planet cores of} $> 14 M_{\oplus}$ {\it that form within observed
protostellar disk lifetimes and are subcritical with respect to the
surrounding gas density.} An ice giant formation theory should also
account for the planets' bulk composition, particularly their 20--50
$\times$ solar tropospheric C/H ratios (Encrenaz 2005). Treating feeding
zone location as a free parameter creates the further challenge of
moving Uranus and Neptune into their current orbits.

Two previous theories attempted to explain both the timely formation and
subsequent orbital evolution of the ice giants. Thommes et al.\ (1999,
2002) proposed that Uranus and Neptune are failed gas giants that formed
between Jupiter and Saturn. Jupiter scattered the ice giants into orbits
with semimajor axes $a > 15$~AU once it reached runaway gas accretion,
while interactions with planetesimals further forced the ice giants
slowly outward. The ``collisional damping scenario'' was put forth by
Goldreich et al.\ (2004a, 2004b). According to Goldreich et al., Uranus
and Neptune formed {\it in situ} from a dynamically cold planetesimal
disk that also produced three other proto-ice giants. The protoplanets
formed quickly despite long dynamical times because the planetesimal
disk scale height fit within the Hill sphere (the protoplanet's zone of
gravitational dominance), leading to high solid accretion rates.
Dynamical friction could no longer damp the eccentricities of the $\sim
5$ trans-Saturnian oligarchs once they attained a surface density
comparable to the surrounding planetesimal disk. The oligarchs suffered
close encounters and the resulting instability ejected all proto-ice
giants but Uranus and Neptune.

%Unfortunately, neither theory presents a complete, incontrovertible
%pathway from solar nebula to present solar system.

The assumptions underlying the order-of-magnitude analysis in Goldreich
et al. (2004a, 2004b) have ultimately proven unreliable.
\cite{levison07} demonstrated that the collisional damping scenario
cannot reproduce the current Solar System: rather than ejecting three of
five ice giants, the trans-Saturnian protoplanets simply spread out and
all planets were retained. Furthermore, the collisional damping scenario
requires that oligarchs grow while planetesimals fragment to sizes $\ll
1$~km. Since low-velocity particles ($v < 10$~cm~s$^{-1}$) in the
COLLIDE-2 microgravity experiment burrowed into the target material
without producing ejecta (Colwell 2003), there is no reason
planetesimals should fragment in the dynamically cold planetesimal disk
required to produce Uranus and Neptune {\it in situ}.

%The Goldreich et al. (2004a, 2004b)
%conclusions relied on order-of-magnitude analysis that Levison \&
%Morbidelli's detailed numerical simulations showed to be incomplete.

The Thommes et al.\ (1999, 2002) ``failed gas giant'' model has
substantial success reproducing the current Solar System and does not
require finely tuned planetesimal behavior. Studies of planet formation
in the 5-10~AU region demonstrate the efficiency of growing ice
giant-sized cores between Jupiter and Saturn [\cite{hubickyj05},
\cite{saturn}]. However, the compositions of Uranus and Neptune strongly
indicate an origin in the trans-Saturnian solar nebula. Tropospheric
abundances of methane show carbon enrichments of 20--50 times solar
(Encrenaz 2005), and interior models find methane mass fractions of
$\sim 20$\% (Marley et al.\ 1995; Podolak et al.\ 1995). The combined
dynamical and chemical model of the solar nebula calculated by
\cite{icelines} shows that the methane condensation front is beyond
Saturn's orbit during the first $5 \times 10^5$ of solar nebula
evolution. Without methane ice present during the planetesimal-building
epoch---which lasts only $5 \times 10^4$ years according to Johansen et
al.\ 2007---neither planet could obtain its methane-rich composition.

%One minor criticism of the failed gas giant model is
%that Neptune's Trojans (2001 QR$_{322}$ and 2004 UT$_{10}$) should not
%have survived a catastrophic scattering off Jupiter or Saturn (Hahn and 
%Malhotra 2005). It is, however, possible that Neptune captured its
%Trojans from heliocentric orbits (Chiang and Lithwick 2005).

The Nice model of planetary dynamics \cite{tsiganis05}, \cite{gomes05},
\cite{morbidelli05} uses initial conditions that place Uranus and
Neptune initially in the methane ice-rich regions beyond 10~AU. In the
Nice model, Neptune and Uranus assume initial semimajor axes of $\sim
12$ and $\sim 15$~to~17~AU. When planetesimal perturbations pull Jupiter
and Saturn across their 1:2 mean motion resonance (MMR), their
eccentricities suddenly increase, forcing close encounters between all
possible pairs of giant planets except Jupiter and Saturn. In about half
of the simulations, Neptune is scattered across Uranus' orbit,
leapfrogging to $\sim 23$~AU within a few $10^5$ years. Slow outward
migration due to interaction with a planetesimal disk pulls Uranus and
Neptune into their current orbits over the course of $\sim 40$~Myr.

%The chemical differences between gas and ice giants add a further
%theoretical hurdle: to  the accreted  With their methane-rich atmospheres and 
%Something about the chemical differences between gas and ice giants here

Although the Nice model explains the current orbits of the giant
planets, it is incomplete without an assessment of the planets' ability
to form in their predicted initial orbits. With high solid surface-
density, methane-rich planetesimals, the protostellar disk model of
\cite{icelines} contains promising initial conditions for ice giant
formation between 12 and 15~AU. None of the three dynamical theories
discussed---the Nice model, the failed gas giant theory and the
collisional damping scenario---treats the growth of the ice giants'
envelopes, a gap in the literature that this work is partly intended to
fill. Verifying that $\sim 10-15 M_{\oplus}$ solid cores can form is an
important step---one which N-body simulations show is extremely
difficult even in the inner nebula (McNeil et al.\ 2005; Chambers
2008)---but one also has to verify that the ice giant atmospheres stay
under $10\%$ of the total planet mass and do not experience runaway
growth.

In this paper, we demonstrate that Uranus and Neptune can form by core
accretion, in the feeding zone approximation, in the trans-Saturnian
solar nebula using the Nice model initial semimajor axis distribution.
In \S \ref{simulation} we describe the numerical methods used in our
experiments. In \S \ref{results} we discuss the results of our core
accretion simulations, focusing on formation timescale, accretion
efficiency and solid/gas ratio. In \S \ref{discussion} we discuss the
strengths and weaknesses of our model as a realistic descriptor of
Uranus and Neptune's formation. We present our conclusions in \S
\ref{conclusions}.

\section{Core Accretion Model}
\label{simulation}

The contraction and buildup of protoplanetary cores and their gaseous
envelopes embedded in our model evolving disk are computed with a
Henyey-type code (Henyey et al.\ 1964), which solves the standard
equations of stellar structure for the envelope. A detailed description
of the core accretion-gas capture code is available in \cite{pollack96},
\cite{bodenheimer00} and \cite{hubickyj05}. Here we explain the initial
conditions used for our experiments and give an overview of our
numerical method, describing its strengths and weaknesses.

We use a core accretion rate of the form
\begin{equation}
{dM_{\rm core} \over dt} = C_1 \pi \Sigma_{\rm solid} R_c R_h \Omega
\label{coregrowth}
\end{equation}
(Papaloizou and Terquem 1999), where $\Sigma_{\rm solid}$ is the surface
density of solid material in the disk, $\Omega$ is the orbital frequency
at the position of the planet, $R_c$ is the effective capture radius of
the protoplanet for solid particles, $R_h = a[M_{\rm planet} / (3
M_*)]^{1/3}$ is the tidal radius of the protoplanet (where $a$ is the
semimajor axis of the protoplanet's orbit), and $C_1$ is a constant near
unity.

Our numerical experiments are based on the feeding zone approximation,
in which the growing embryo accretes planetesimals from an annulus
extending $\sim 4 R_h$ on either side of its semimajor axis (Kary \&
Lissauer 1994). Planetesimals in the feeding zone exit only by being
accreted; they are not scattered out of the system, and they encounter
the core at the Hill velocity, $v_h = R_h \Omega$. As the protoplanet
grows, its tidal radius expands to include previously undepleted regions
of the solar nebula. New planetesimals enter the feeding zone as it
expands, and the solid surface density evolves according to the
competing effects of planetesimal loss by accretion and gain by feeding
zone expansion. The fact that the feeding zone model does not include
planetesimal loss by scattering limits its accuracy in the
trans-Saturnian region, where planetesimals may be stirred to velocities
near the local escape speed of the Solar System (Levison \& Stewart
2001; Levison \& Morbidelli 2007).

We view our core accretion experiments as a ``first pass'' at
reconciling the Nice model with planet formation theory. Our simulations
will tend to overestimate the solid accretion rate, so if we cannot form
Uranus and Neptune on a reasonable timescale, the initial conditions in
the Nice model cannot possibly represent Uranus and Neptune's formation
zones. If we can successfully produce planets with the proper ice giant
mass and composition, the detailed planetesimal dynamics at 12--15~AU
will warrant further investigation with N-body simulations. Simulations
such as those of Chambers (2008) can establish whether or not the
initial $1 M_{\oplus}$ seed core can form and resist orbital decay
between 12 and 15~AU, and whether our low-velocity planetesimals are an
adequate approximation.

Once the protoplanet reaches $\sim 5 M_{\oplus}$ it begins to accrete
gas at the combined rate at which (1) its Bondi sphere is evacuated due
to cooling and contraction, and (2) its Bondi sphere expands due to
increasing mass. Gas flows freely from the nebula into the evacuated
volume. The outer boundary conditions of the gaseous atmosphere include
the decrease with time in the background nebular density and
temperature, modeled by \cite{icelines}. Since ice giants are not
massive enough to limit their gas accretion by opening a gap in the
gaseous disk, we carry the simulations beyond Uranus and Neptune's
present masses in order to assess the likelihood that they remain
low-mass ice giants. Since grain settling in the protoplanetary envelope
reduces envelope opacity where grains exist (Podolak 2003), we adopt
grain opacities of $\sim 2\%$ of the interstellar values used in
\cite{pollack96}. However, \cite{saturn} demonstrated that grain opacity
has little effect on formation timescale and planet composition in the
trans-Saturnian region of the solar nebula.

Since the ice giants swap orbits in $\sim 50\%$ of the Nice model
simulations, we generically simulate the formation of two ice giants in
trans-Saturnian orbits without identifying which planet embryo is
proto-Uranus and which is proto-Neptune. In accordance with the initial
conditions for the Nice model, we place ``Planet I'', the innermost ice
giant during the formation epoch, at 12~AU. ``Planet O'', the outermost
ice giant, begins forming at 15~AU. Which embryo eventually becomes
which planet depends on whether or not the ice giants swap orbits.

Solid durface densities available for planet formation are taken from
\cite{icelines} and are based on a detailed chemical inventory of 211
gas and ice species. To allow for the time needed to build the $1
M_{\oplus}$ seed core and the planetesimals, we adopt a 0.15~Myr offset
between the initial formation of the solar nebula and the beginning of
planet formation (Dodson-Robinson et al.\ 2008). Planetesimal formation
models based on the streaming instability show that the timescale can be
as short as $5 \times 10^4$ years and is not a strong function of
location in the disk (Youdin and Goodman 2005; Johansen et al.\ 2007),
so our 0.15~Myr planetesimal and seed embryo formation timescale is
realistic. Planet I at 12~AU begins forming in a feeding zone with
8.4~g~cm$^{-2}$ of solids while Planet O's feeding zone at 15~AU
contains 6.4~g~cm$^{-2}$ of solids. For comparison, \cite{pollack96}
assumed only 0.75~g~cm$^{-2}$ of planetesimals were available in Uranus'
{\it in situ} feeding zone at 19~AU.

Since the goal of our simulations is to bridge the initial solar nebula
inventory and the final planet composition---both bulk gas/solid ratio
and measured atomic abundances---our numerical experiments are designed
to be as deterministic as possible. The three ``free parameters'' in our
core accretion model are semimajor axis, solid surface density of
planetesimals, and atmospheric opacity. Initial planet semimajor axes
are taken from the Nice model of planetary dynamics so as to be
consistent with late-stage orbital evolution; solid surface densities
come from a detailed chemical model whose few free parameters were
constrained by observational and laboratory data; and Dodson-Robinson et
al.\ (2008) convincingly demonstrated that atmospheric opacity has
little effect on planet composition or formation timescale in the outer
solar nebula, where isolation masses are high. Here we are testing
whether or not one single disk model (Dodson-Robinson et al.\ 2009) can
serve as a formation platform for all four giant planets. If it can, we
will have made substantial progress toward determining the physical
conditions in the true solar nebula.

%Captured
%planetesimals may be ablated by the massive gaseous envelope during the
%late stages of planet growth. If a planetesimal deposits 50\% or more
%of its mass in the envelope, it is assumed captured, and the resulting
%debris sinks to the core. 

\section{Results}
\label{results}

Figure 1 shows the result of our core accretion simulations. Planet I
reaches its present mass in 3.8~Myr while Planet O requires 5.3~Myr to
form. However, there is enough solid material remaining in the nebula to
allow the ice giants to continue growing beyond $17.2 M_{\oplus}$
(Neptune) and $14.5 M_{\oplus}$ (Uranus). One characteristic of planet
formation in solid-rich feeding zones is that the solid cores do not
reach isolation mass (Dodson-Robinson and Bodenheimer 2009). In the
solid-rich solar nebula of \cite{icelines}, Planet I's isolation mass is
$88 M_{\oplus}$ while Planet O's is $167 M_{\oplus}$. Previous
simulations of forming planets that don't reach isolation mass
(Dodson-Robinson et al.\ 2008, Dodson-Robinson and Bodenheimer 2009)
show that $\dot{M}$ monotonically increases with time. The canonical
Phase II, in which the planet's growth is limited by the rate at which
gas can enter the Hill sphere (Pollack et al.\ 1996, Hubickyj et al.\
2005), does not occur.

Upon attaining their current masses, both planet cores are still
subcritical and have yet to trigger rapid gas accretion. Here our
results agree with those of Rafikov (2006), who showed that critical
core mass is not simply a constant $10 M_{\oplus}$ (e.g.\ Mizuno 1980),
but is a strong function of planetesimal accretion rate, semimajor axis
and gas density. In our simulations, the continued fast accretion of
planetesimals deposits kinetic energy that stabilizes the ice giant
atmospheres against collapse even as the planet masses reach $15
M_{\oplus}$.
%The luminosity provided by solid accretion would only be
%distance-independent if all cores, everywhere in the nebula, reached
%isolation mass before accreting gas.
In the trans-Saturnian nebula,
dynamical times are so long and isolation masses are so large ($M_{\rm
iso} \propto a^3$) that is not possible for any protoplanet to cleanly
sweep up the planetesimals in its feeding zone.

If allowed to grow beyond its present mass, Planet I would
enter rapid gas accretion at 4.3~Myr with a total mass of $43
M_{\oplus}$. At 5.7~Myr, Planet O had not yet begun hydrodynamic gas
accretion. We therefore cannot evaluate the critical core mass in the
feeding zone at 15~AU, but suspect it is somewhat higher than on Planet
I's orbit at 12~AU because of the lower nebular gas density. We
calculate that both planets contain 7\% hydrogen and helium gas by mass
and 93\% heavy elements, in agreement with bulk compositions inferred
from rotation rates (Podolak and Reynolds 1981, Guillot 2005).

%The fact that Uranus and Neptune reach $\sim 15 M_{\oplus}$ without
%undergoing hydrodynamic accretion may be surprising in the context of
%the traditional core accretion theory, in which the critical core mass
%of 5--15~$M_{\oplus}$ does not vary with location in the nebula (Mizuno
%1980; Stevenson 1982). However, subsequent analytical work by
%\cite{rafikov06} demonstrates that critical core mass is a strong
%function of nebular gas density---denser regions can be destabilized by
%smaller cores---and solid accretion rate onto the core. 

Is it realistic that all but 10--15~$M_{\oplus}$ of planetesimals are
lost from each feeding zone, without forming planet cores? If Uranus and
Neptune are to remain at their present masses, the
planetesimal-to-planet conversion efficiency in this model is extremely
low at 17\% for Planet I and 9\% for Planet O. Fortunately, we can
perform a sanity check on the planet formation efficiency by calculating
the accretion-to-ejection probability ratio in the strongly scattering
regime, where the protoplanet stirs planetesimals to of order the local
solar system escape velocity. N-body simulations by Levison \& Stewart
(2001) and Levison \& Morbidelli (2007) show that the strong-scattering
assumption is appropriate for the trans-Saturnian solar nebula. Ida \&
Lin (2004) demonstrate that the accretion-to-ejection probability ratio
of strongly scattered planetesimals is
\begin{equation}
f_{\rm cap} \approx \left ( \frac{2 G M_{\odot} / a}{G M_{\rm core} / R_{\rm core}} \right )^2 ,
\label{fcap}
\end{equation}
where the denominator is the square of the typical planetesimal velocity
with respect to the core and the numerator is the square of the local
escape velocity from the sun. In Eq.\ \ref{fcap}, $R_{\rm core}$ is the
physical radius of the core, not a gravitationally enhanced capture
radius. The probability that a given planetesimal will be scattered,
rather than accreted, is therefore
\begin{equation}
P_{\rm sca} \approx \frac{1}{f_{\rm cap} + 1}.
\label{psca}
\end{equation}

For a core density of 3~g~cm$^{-3}$ and mass $15 M_{\oplus}$, $f_{\rm
cap} \approx 0.20$ and $P_{\rm sca} \approx 0.83$ at 12~AU. At 15~AU,
$f_{\rm cap} \approx 0.14$ and $P_{\rm sca} \approx 0.88$.  Once
reaching ice giant mass, growing embryos can accrete less than $1/5$ of
the available planetesimals and should reach a stage of self-limited
solid accretion. Scattering therefore limits the core masses of Uranus
and Neptune and efficiently clears away the remaining planetesimals.

%The protoplanet core asymptotically approaches a maximum mass of
%\begin{equation}
%M_{\rm core} \sim 250 \left ( \frac{\rho_{\rm core}}{1 {\rm g;cm^{-3}}}
%\right )^{-1/2} f_{\rm cap}^{-3/4} \left ( \frac{a}{\rm 1;AU} \right
%)^{-3/2} M_{\oplus} .
%\label{mcoremax}
%\end{equation}

%\cite{ikoma01} note that convective protoplanetary envelopes arise in
%dense, cool disks. The solid-rich solar nebula is such a disk: its total
%5mass is 0.12~$M_{\odot}$ and it cools to $< 35$~K at the midplane beyond
%12~AU.

% However, the intial critical mass estimates
%assumed radiative protoplanetary atmospheres. \cite{wuchterl93} first
%noticed that for convective atmospheres, the nebular boundary conditions
%do control the critical mass.

\section{Discussion: Model strengths and weaknesses}
\label{discussion}

%The fact that Neptune is more massive
%than Uranus is a good argument that it formed interior to Uranus, as
%predicted by the Nice model: given the fact that mass accretion rate
%scales as Keplerian speed (Eq. 1), one expects final core mass to
%decrease with feeding zone radius.

We have demonstrated that timely formation of Uranus and Neptune in the
trans-Saturnian solar nebula, within the $\sim 5$~Myr disk lifetimes
found by young cluster observations and photoevaporative models of
T-Tauri disks (Alexander et al.\ 2006; Currie et al.\ 2009) and without
extreme planetesimal eccentricity damping (Goldreich et al.\ 2004a,
2004b), is possible in the feeding zone approximation. In this section
we assess the extent to which our core accretion simulations, which use
the initial semimajor axes of the Nice model and the solid-rich disk of
\cite{icelines}, provide a realistic picture of the formation of Uranus
and Neptune. Our model's strengths include the ice giants' predicted
compositions and the fact that both are comfortably below critical mass.
Our model's weaknesses are lack of coevality---the outer ice giant O
requires 1.5~Myr more for formation than the inner ice giant I---and the
simplified description of planetesimal accretion provided by the feeding
zone approximation.

\subsection{Weakness: Ice giant coevality}

Although our cores stay subcritical and we do not need to cut off
hydrodynamic gas accretion to replicate Uranus and Neptune's gas-poor
composition, both planets reach their present masses in the midst of
rapid solid accretion, well past the ``elbow'' of the $M(t)$ curve
(Fig.\ 1). Why should they not continue to grow into massive cores
analogous to HD 149026b (Sato et al.\ 2005)? From Fig.\ 1, we see that
Planet I reaches $70 M_{\oplus}$ at the end of the simulation. In the
feeding zone approximation, there is nothing to stop it growing to
Saturn's mass or greater while Planet O slowly moves toward $15
M_{\oplus}$.

The simplest explanation is self-limiting accretion due to planetesimal
scattering, as discussed in \S \ref{results}. Planet I would reach $\sim
15 M_{\oplus}$, begin to scatter far more planetesimals than it
accretes, and dramatically slow its core growth. Planet O would
similarly halt its growth in due course. If a self-regulating mechanism
impedes core growth in the trans-Saturnian nebula, there may be no need
for Planet I and Planet O to be coeval. Based on an
accretion-to-ejection probability ratio of 0.1 (Lin \& Ida 1997), Ida \&
Lin (2004) derive a maximum planet core mass of
\begin{equation}
M_{\rm core} \approx 1.4 \times 10^3 \left ( \frac{a}{\rm 1 \; AU}
\right )^{-3/2} M_{\oplus} .
\label{maxcore}
\end{equation}
Planet I's core could not grow beyond $34 M_{\oplus}$ while Planet O's
core would be limited to $24 M_{\oplus}$. Furthermore, as demonstrated
in \S \ref{results}, the accretion rate reduction kicks in before the
cores reach their absolute maximum mass. Strong scattering of
planetesimals by the growing protoplanets could hold off rapid solid
growth long enough for Planet O's mass to catch up with Planet I's.

To demonstrate the effects of planetesimal scattering on ice giant
composition and growth rate, we simulated the growth of both planets,
this time modifying the accretion rate to take into account planetesimal
scattering. We first calculated the solid growth rate $\dot{M}$
according to Equation \ref{coregrowth}. Then, since $f_{\rm cap}$
(Equation \ref{fcap}) is defined as the ratio of the probabilities of
accretion and ejection for a given planetesimal, it follows that
planetesimals are ejected from the feeding zone at a rate of
\begin{equation}
\dot{M}_{\rm eject} = \frac{\dot{M}}{f_{\rm cap}}.
\label{ejection}
\end{equation}
At each timestep, we subtracted the sum of the accreted and the ejected
planetesimal mass, $\dot{M} \Delta t + \dot{M}_{\rm eject} \Delta t$,
from the total solid feeding zone mass.  Figure 2 shows the results of
the simulations that include planetesimal ejection. As expected, both
ice giants take longer to reach their current masses when their solid
accretion is slowed by scattering: Planet I forms in 4.0~Myr and Planet
O forms in 6~Myr.

With self-limiting accretion, however, we risk losing our solid-rich
composition. Continual planetesimal accretion is what stabilizes the ice
giant atmospheres against hydrodynamic collapse. As demonstrated by
Hubickyj et al.\ (2005), suddenly cutting off accretion once a planet
core reaches $10 M_{\oplus}$ {\it hastens} the onset of rapid gas
accretion. We see this behavior in the left panel of Figure 2 as Planet
I, within a few $10^5$~years after reaching ice giant mass, begins to
accrete gas at an exponential rate. Since Planet I begins to halt clear
its feeding zone and slow its solid accretion before Planet O finishes
forming, it evolves toward Jupiter's mass and composition. Even where
planetesimal scattering is efficient, we still require a way of halting
the innermost ice giant's growth. Strong planetesimal scattering does
not mitigate the need for Uranus and Neptune to form on similar
timescales.

For Planet O, with its formation timescale of 6~Myr coming uncomfortably
close to the upper limit of protostellar disk lifetimes (Haisch et al.\
2001, Alexander et al.\ 2006), it is sensible to speculate that its
growth was halted by the dissipation of the solar nebula. Even though
Planet O is primarily accreting solids, rather than gas, between 5 and
6~Myr, nebular gas provides eccentricity damping to the planetesimals,
helping keep them in the feeding zone. Without the gas, the planetesimal
velocity dispersion rapidly increases, cutting off solid accretion
except in the case of stochastic, high-velocity collisions (e.g.
Fernandez and Ip 1984). Such a collision could be responsible for
Uranus' extreme rotation axis tilt. Postulating the dissipation of the
Nebula as the mechanism for halting Planet O's solid growth, however,
does nothing to rectify the coevality problem.

%In Neptune's case, postulating solar nebula dissipation as the mechanism
%for solid growth cutoff has the dire consequence that Uranus will not
%grow beyond $6 M_{\oplus}$. Either the initial formation distances
%predicted by the Nice model are incorrect, or some other process has to
%halt Neptune's growth.

%It is therefore tempting to return to the failed
%gas giant model of Thommes et al. (1999; 2002). However, we caution that
%it is nearly impossible to form two planets whose masses are so similar
%as Uranus and Neptune's without placing them in the {\it same} feeding
%zone, where they would certainly interact. The fact that solid accretion
%rate is proportional to Keplerian speed means that where planet growth
%is cut off suddenly---as is the case with photoevaporation, where disk
%dispersal timescales are $\sim 10^5$ years (Alexander et al. 2005)---the
%inner ice giants will always be more massive than the outer ones. (It is
%less clear whether this mass-distance relationship should hold in the
%case of giant planets, since they accrete gas so quickly in the final
%stages of formation. Since gap opening slows gas accretion, the final
%mass-distance relationship may reflect the disk aspect ratio.)

A second possibility for keeping both ice giants at the same mass is
that Planet I was scattered to a wide orbit---with correspondingly long
dynamical time and low surface density---during the dynamical
instability of the four outer planets. In roughly half of the Nice model
simulations, Neptune and Uranus cross orbits, with Neptune
(corresponding to Planet I in this scenario) moving from $\sim 12$ to
$\sim 23$~AU in about $10^5$ years.  Given the long orbital timescale at
23~AU (110 years vs. 41 years at 12 AU), Neptune's accretion would be
nearly cut off, especially in light of its subsequent outward migration
due to planetesimal scattering.  Although the authors of the Nice model
use the Jupiter--Saturn 1:2 MMR crossing to explain the Late Heavy
Bombardment occurring 600~Myr after the birth of the solar system (Gomes
et al.\ 2005), this resonance crossing could have occurred at any time
(or multiple times) during the evolution of the solar nebula. If the
Jupiter-Saturn 1:2 MMR crossing caused the ice giants to remain at
similar masses by halting Planet I's accretion, Neptune must have formed
interior to Uranus.

Within the feeding zone approximation, there is no entirely satisfying
way to explain why Uranus and Neptune have nearly the same mass. More
detailed N-body simulations are needed to investigate the degree to
which the ice giants limit their own accretion in the 12--15~AU region
of the solar nebula. 

\subsection{Strength: Ice giant composition}
\label{composition}

Although traditional core accretion-gas capture simulations cannot
entirely explain the coeval formation of the ice giants without
requiring an external growth-limiting mechanism, they still provide
strong constraints on the conditions required to produce ice giants.
In this section, we discuss the characteristics of feeding zones
that can produce solid-rich, gas-poor planets.

In Section \ref{results} we showed that by keeping Uranus and Neptune's
cores subcritical, we reproduce their solid-rich compositions of $> 90
\%$ solid by mass (Podolak and Reynolds 1981) without needing to halt
gas accretion during the extremely short hydrodynamic growth phase. We
also bypass the long opacity-limited gas contraction phase (Phase II),
which lasted $\sim 4$~Myr in the simulations of \cite{pollack96},
because our ice giants never reach isolation mass. Protoplanetary
envelope opacity should not substantially affect the formation timescale
or bulk composition of Uranus or Neptune (Dodson-Robinson et al.\ 2008).
The solid/gas ratios we calculate are therefore insensitive to the
composition and size of grains in the ice giant envelopes.

Figures 2 suggests that continual planetesimal accretion is required to
limit the mass of the protoplanet atmosphere. Rapid gas accretion begins
as soon as Planet I's solid growth rate begins to level off, at the
inflection point of the $M_{\rm core}(t)$ curve. The more massive the
protoplanet, the more thermal energy is required to stabilize its
atmosphere. Our results suggest that {\it once a protoplanet core grows
beyond the fiducial critical core mass of} $10 M_{\oplus},$ {\it it will
trigger runaway gas accretion unless its solid accretion rate keeps
increasing with time.} Indeed, we conducted one experiment in which
solid accretion was suddenly cut off after 3.5~Myr of planet growth with
a core mass of $12 M_{\oplus}$. As expected from the results of Hubickyj
et al.\ (2005), the onset of rapid gas accretion was immediate.

To keep the solid/gas mass ratio high and ensure the planetary
composition does not become Jupiter-like, a large supply of available
planetesimals is critical. Planetesimals could either come from the
initial feeding zone or be accreted as the protoplanet migrates into
undepleted regions of the disk (Alibert et al. 2005). We tested the
effect of planetesimal availability on final planet composition by
varying the initial solid surface density in the feeding zones and
calculating the solid/gas ratio once the protoplanet reached ice giant
mass ($14.5 M_{\oplus}$ for Uranus and $17.2 M_{\oplus}$ for Neptune).
These tests included planetesimal scattering in the accretion rate
calculation. High values of initial planetesimal surface density
$\Sigma_0$ ensure that either the feeding zone will never be depleted of
planetesimals, or in the case of migration, mimic a protoplanet always
moving into areas of the solar nebula with a fresh planetesimal supply.

Figure 3 shows solid/gas ratio as a function of $\Sigma_0$. Since we do
not know which ice giant formed in which feeding zone, we calculate
pairs of solid/gas ratios for each initial value of $\Sigma_0$, one
corresponding to the final mass of Uranus and the other to the final
mass of Neptune. For both feeding zones, at distances of 12 and 15~AU,
solid/gas ratios within 25\% of the fiducial ice giant composition of
$\sim 90\%$ solids can only be reached for a narrow range of initial
surface densities between 6 and 11~g~cm$^{-2}$. Note that while Pollack
et al.\ (1996) reproduced Uranus' composition by requiring the planet to
form {\it in situ} in a 19~AU feeding zone with only $0.75$~g~cm$^{-2}$
of planetesimals, its formation timescale was 16~Myr---far longer than
observed protostellar disk lifetimes. For surface densities from 6 to
11~g~cm$^{-2}$, we find formation timescales of 8.8--2.9~Myr at 12~AU
and from 9.5--3.0~Myr at 15~AU. {\it Solid-rich feeding zones with} $6 <
\Sigma_0 < 11$~g~cm$^{-2}$ {\it at positions consistent with the Nice
model can give both appropriate formation timescales and bona fide ice
giant compositions.}

Even though we require that planetesimals always be present in the ice
giant feeding zones as long as the gas disk lasts, our work does not
necessarily conflict with the Nice model results. Gomes et al.\ (2005)
invoke a sharp inner edge of the planetesimal distribution at 15.3~AU in
order to delay the onset of Jupiter and Saturn's 1:2 MMR crossing and
allow the resulting dynamical shakeup to coincide with the Late Heavy
Bombardment 700~Myr after the birth of the solar system. Their Figure 1b
shows that the time of the 1:2 MMR crossing is a strong function of the
location of the inner disk edge: the greater the distance between
Jupiter and Saturn and the planetesimals that perturb their orbits, the
smaller the perturbations to the gas giant orbits and the longer it
takes to force them through the resonance. Gomes et al.\ chose 15.3~AU
because planetesimals outside that radius have a dynamical lifetime,
defined as the time required for the planetesimal to pass within the
Hill sphere of a planet, longer than their assumed gas disk lifetime of
3~Myr.

However, in order to build the planets, the dynamical lifetime of
planetesimals in their feeding zones cannot possibly be longer than the
nebula lifetime---otherwise, no protoplanet could ever grow. For
a planetesimal disk with an inner edge at 12~AU, coinciding with our
innermost ice giant, we see from Figure 1 of Gomes et al.\ (2005) that
the time till Jupiter and Saturn's 1:2 MMR crossing is $\sim 5$~Myr.
Tsiganis et al.\ (2005) found a 6~Myr lag time between the beginning of
their simulations and the 1:2 MMR crossing. Our simulations demonstrate
that the orbital evolution of the giant planets must take place shortly
after Uranus and Neptune finish forming. Sculpting the planetesimal
distribution to allow the giant planets' orbital evolution to trigger
the Late Heavy Bombardment does not allow Uranus and Neptune to retain
their solid-rich composition.

In the \cite{icelines} solid-rich disk model, the methane condensation
front is at 10~AU at the beginning of planet formation. Any
planetesimals beyond 10~AU contain 10\% methane ice by mass.
\cite{podolak95} calculated a series of ice giant interior models using
three concentric layers: a rock+metal core, an ice shell with H$_2$O,
CH$_4$, NH$_3$ and H$_2$S in solar proportions, and an atmosphere of
H$_2$, He and small amounts of the aforementioned ices.  For both
planets, the models that best reproduced the gravitational moments
measured by {\it Voyager} had ice shells of $\sim 13 M_{\oplus}$.
Podolak et al.\ found that total methane mass fraction in the ice giants
is $\sim 20$\%, whereas our simulations construct planets that are $\sim
9$\% methane by mass. The reason for this discrepancy is that we include
refractory CHON material in the planet cores, while Podolak et al.\
assume iron and silicate cores.

We calculate total carbon mass fractions of $22 \%$, in agreement with
those derived by \cite{podolak95} and \cite{marley95}. {\it We have thus
created the first planet formation model to reproduce the observed
carbon enrichment of both ice giants (Encrenaz 2005).} We predict that
the methane present on Triton's surface (e.g.\ Hicks and Buratti 2004)
and in the ice giant atmospheres was accreted as ice from the primordial
solar nebula. The Saturn formation model of Dodson-Robinson et al.\
(2008) came to the same conclusion about the ammonia ice present in
Saturnian satellites (Prentice 2007; Freeman et al.\ 2007).

The fact that we have reproduced Uranus and Neptune's bulk composition
and carbon enrichment so closely places an important constraint on their
feeding zone location: the presence of methane ice in planetesimals is
required during the ice giant formation epoch. Based on both chemically
evolving models of the solar nebula (Dodson-Robinson et al.\ 2009) and
the relatively carbon-poor compositions of Jupiter and Saturn (Encrenaz
2005)---which indicate that they did not form in methane ice-rich
regions---{\it we can reject any formation model in which the ice giants
form inward of Saturn's orbit.}

\section{Conclusions}
\label{conclusions}

We have shown that it is possible to assemble Uranus and Neptune from $1
M_{\oplus}$ seed cores in a $0.12 M_{\odot}$ solar nebula, with initial
semimajor axes of 12~AU and 15~AU as predicted by the Nice model (Gomes
et al.\ 2005; Morbidelli et al.\ 2005; Tsiganis et al.\ 2005).  The
resulting growth timescales are 3.8~Myr for the innermost ice giant and
5.3~Myr for the outermost ice giant.  However, numerous details still
remain to be fully resolved:

\begin{enumerate}

\item What is the migration history of the planets after formation and
can we explain the current orbits?

\item If Planets I and O do not change places, how can Planet I's lack
of a massive gaseous atmosphere be explained?

\item If Planets I (Neptune) is scattered to a wide orbit while the disk
is still gas-rich, what happens to Planet O (Uranus)?

\item How can the Late Heavy Bombardment be explained if the giant
planets' orbital evolution must take place soon after they finish
forming?

\end{enumerate}

Because we chose the initial semimajor axes of the Nice model, both ice
giants' post-formation orbital evolution may be explained according to
that model. If Uranus and Neptune form between 12 and 15~AU in the
feeding zone approximation, they end up with either different formation
timescales or different masses. If the proto-ice giants begin limiting
their own solid accretion rates by ejecting planetesimals, the outermost
planet core can continue to grow after the innermost core has assembled
the bulk of its mass, but the inner planet may evolve into a gas giant.
One way to form Uranus and Neptune at 12~AU and 15~AU but keep their
near-equal masses may be for Neptune (the innermost planet in this
scenario) to get scattered out to a wide orbit upon reaching its present
mass---but in this scenario we cannot be assured of Uranus' survival and
continued growth.

One way our results differ from the scenario described by the Nice model
suite of papers is that the orbital evolution of the giant planets must
take place shortly after Uranus and Neptune finish forming. No disk
where Uranus and Neptune fall in a gap in the planetesimal distribution
{\it at any time during their growth} can form planets with $\sim 90 \%$
solids by mass. As stated in \S \ref{discussion}, we still lack a
mechanism for checking Planet I's growth while Planet O finishes
forming. If Neptune is the innermost ice giant, Jupiter and Saturn's 1:2
MMR crossing might perform such a function by suddenly sending Neptune
into an extremely wide orbit.

The formation timescale of the ice giant forming at 15~AU is quite close
to both the upper limit of observed protostellar disk lifetimes (Haisch
et al.\ 2001; Currie et al.\ 2009) and the predicted onset of
photoevaporation for the solar nebula (Alexander et al.\ 2006),
indicating that quick ice giant formation in the trans-Saturnian nebula
is only marginally possible even in the optimistic feeding zone
approximation. Since the compositions of Uranus and Neptune strongly
indicate that they originated in the trans-Saturnian solar nebula,
however, we encourage more detailed N-body investigations of embryo
growth rates between 12 and 15~AU.

Without knowing for sure whether Uranus and Neptune swapped orbits, we
cannot pair each planet with its unique formation zone. However, the
chemically evolving solar nebula simulations of \cite{icelines} show
that the CO and N$_2$ ice lines fall between 12 and 15~AU. Detailed
measurements of the carbon and nitrogen abundances of both planets and
their satellites could resolve the question of which planet formed
outermost in the solar nebula: the planet that formed near 15~AU should
be slightly more carbon- and nitrogen-rich than the planet that formed
near 12~AU. Such detailed measurements of Uranus and Neptune's
atmospheric compositions await another space mission to the outer
planets, but ground-based spectroscopy of their satellites (e.g.\ Stern
\& McKinnon 2000; Bauer et al.\ 2002) may make some progress on this
question. Our formation model reproduces the ice giants' solid/gas
ratios and carbon abundances inferred from gravitational moment
measurements and spectroscopy.

Our results indicate that planet formation in the outer nebula is a
fundamentally inefficient process. Even in the feeding zone
approximation, where we calculate what is best viewed as an upper limit
to accretion rate, building ice giants within the lifetime of
the solar nebula requires substantially more solid and gas mass than the
planet can ever accrete. The dynamical times in the massive
\cite{icelines} disk (total mass $0.12 M_{\odot}$) beyond 15~AU are
still too long to allow gas or ice giant formation. From the combined
constraints of the methane ice line location and the upper limits to ice
giant growth rates, we can constrain the ice giant formation zone to
between 10 and 15~AU.

%Future N-body simulations that focus on the
%formation of Uranus and Neptune should investigate this region of the
%solar nebula.

The fact that the solar nebula model of \cite{icelines} can form
all the giant planets at locations at or near those predicted by the
Nice model, as well as reproduce important features of the compositions
of the planets and their satellites---such as ammonia enrichment in the
Saturnian system (Dodson-Robinson et al.\ 2008)---indicates that
substantial progress has been made at modeling the ice inventory and
chemical evolution of the solar nebula. Here we have created the first
successful progression (to the best of our knowledge) from 
primordial solar system ice inventory to detailed planet composition.
This research marks a substantial step forward in connecting both the
dynamical and chemical aspects of planet formation. Although we cannot
say that the solid-rich solar nebula model (Dodson-Robinson et al.\
2009) gives {\it exactly} the appropriate initial conditions for planet
formation, rigorous chemical and dynamical tests have at least revealed
it to be a {\it viable} model of the early solar system.

%The only ways for Uranus and Neptune to reach their
%present masses at the same time are (1) if the planetesimal surface
%density in Uranus' feeding zone at 15~AU is {\it higher} than in
%Neptune's at 12~AU, or (2) if the ice giants migrate outward during
%formation. \cite{ward97} demonstrated that Type I migration rates can
%be quite high ($\sim 1$~AU per $10^4$~years), but the predicted
%migration direction is generally inward toward the star.

Support for S.D.R.'s work was provided by NASA through the Spitzer Space
Telescope Fellowship Program. P.B. received support from the NASA
Origins Grant NNX08AH82G. We thank both referees for comments that
improved the clarity and focus of the manuscript.

\begin{figure}
\plottwo{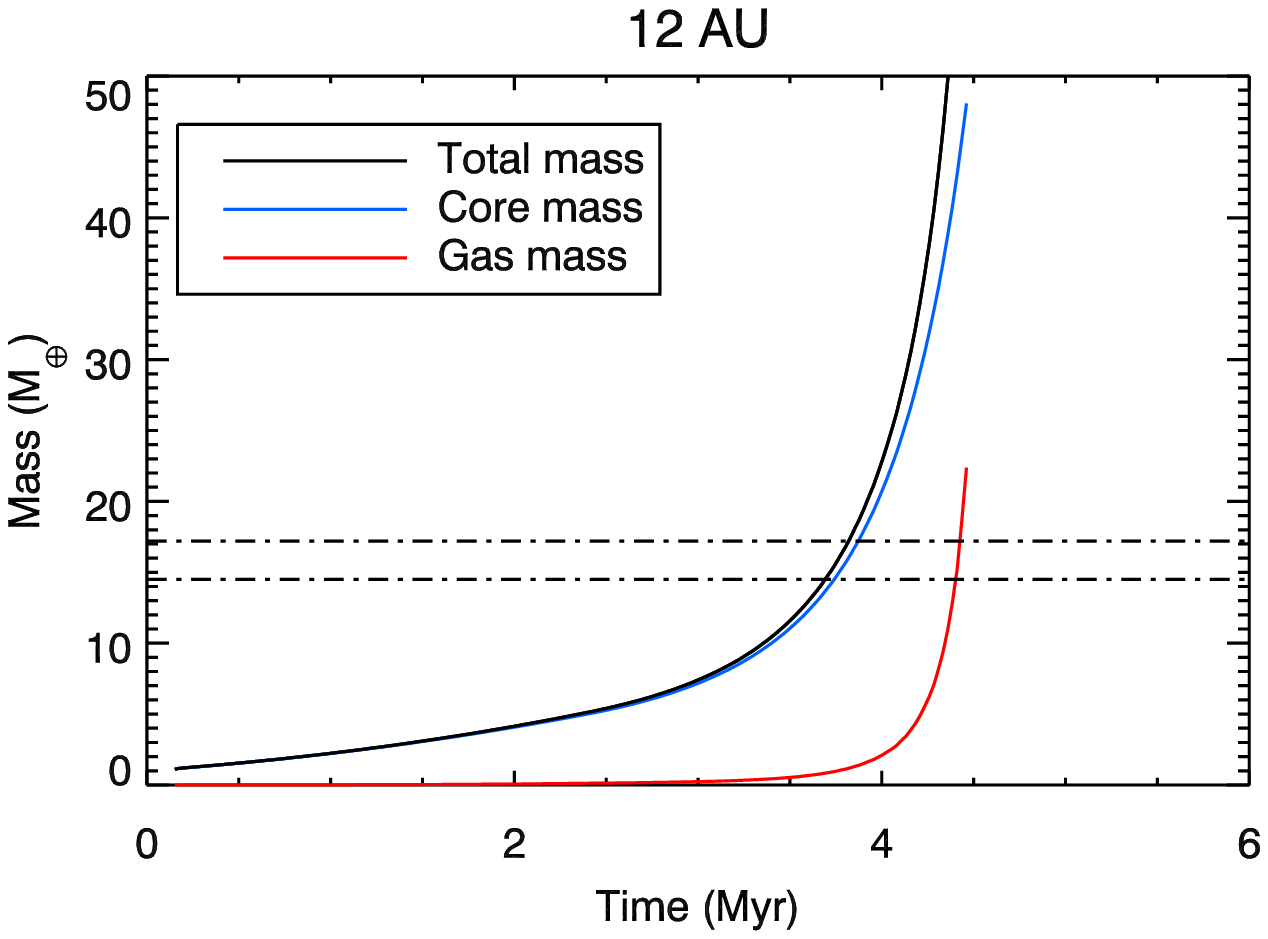}{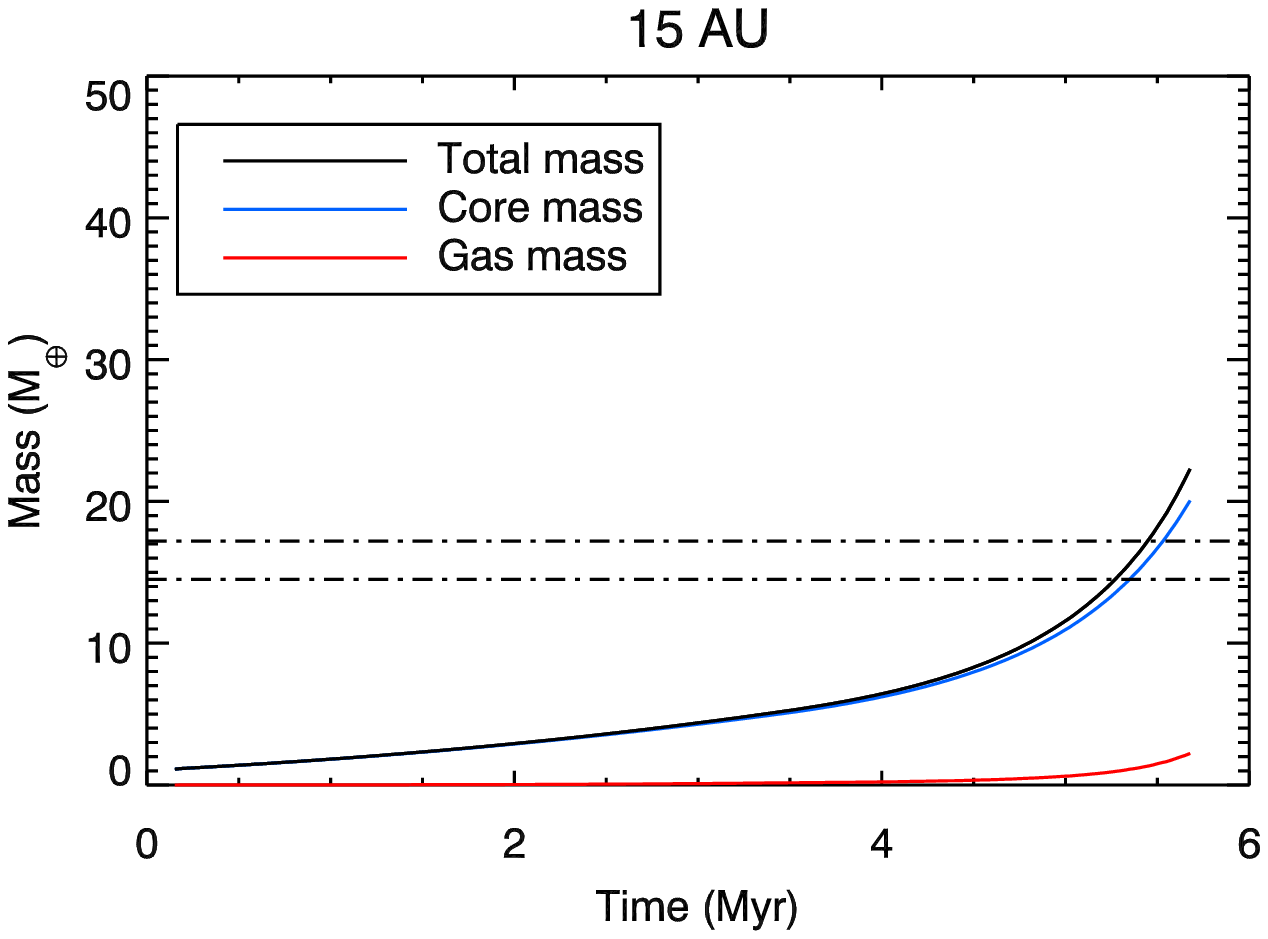}
\caption{
%\clearpage
%{\bf Figure 1 Caption:}
Aided by a solid-rich solar nebula, Uranus and Neptune grow in the
trans-Saturnian region in under 5.5~Myr. Left: Forming at 12~AU
according to the predictions of the Nice model (Tsiganis et al. 2005),
the innermost ice giant attains its current mass in only 3.8~Myr. Right:
At 15 AU, the outer-forming ice giant reaches its present mass in
5.3~Myr. The dashed lines show the planets' current masses. The
solid-rich nebula contains enough planetesimals for the ice giants to
grow well beyond their current sizes, given efficient accretion and
enough time.
}
\label{ungrow}
\end{figure}

\begin{figure}
\plottwo{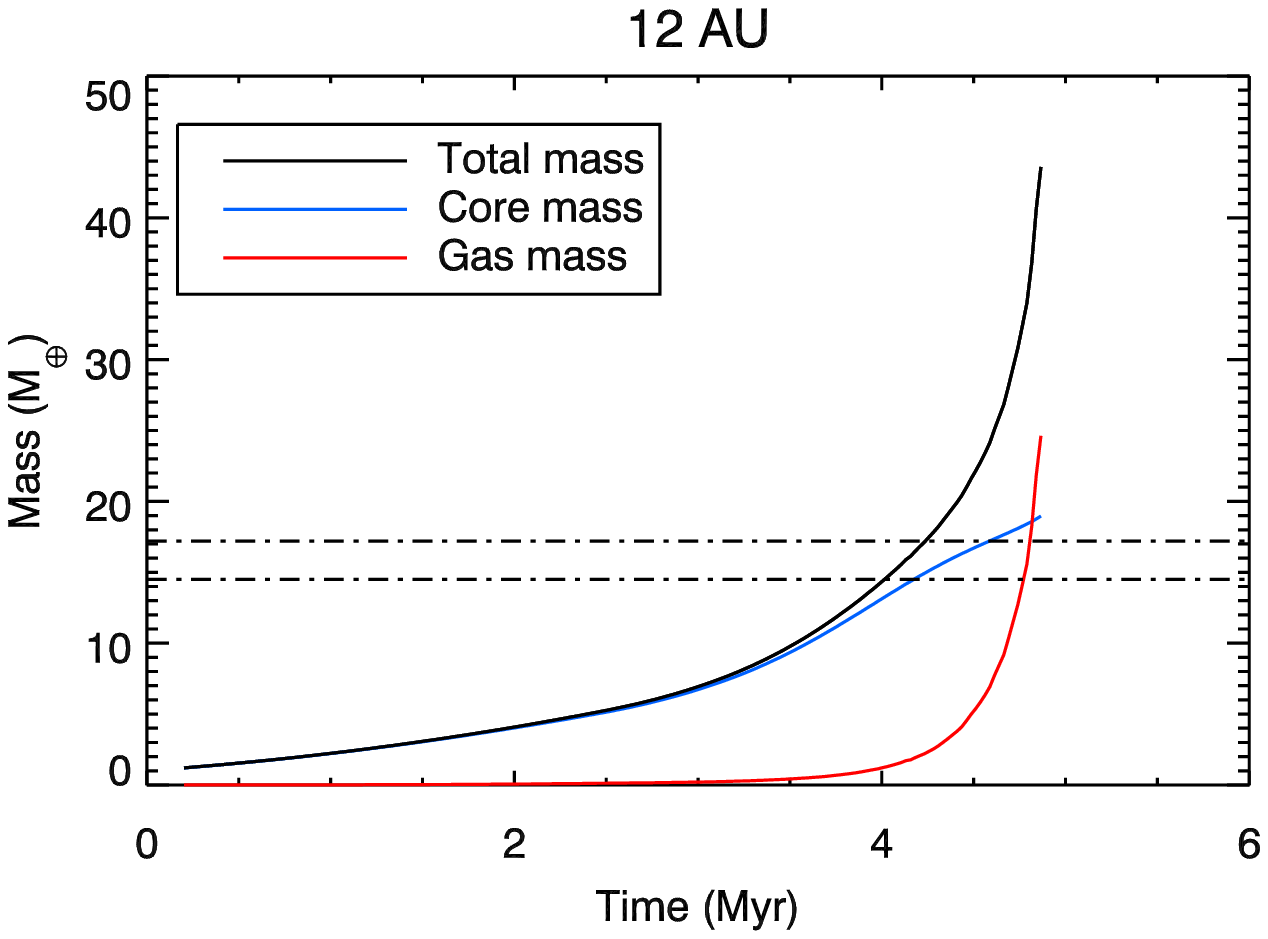}{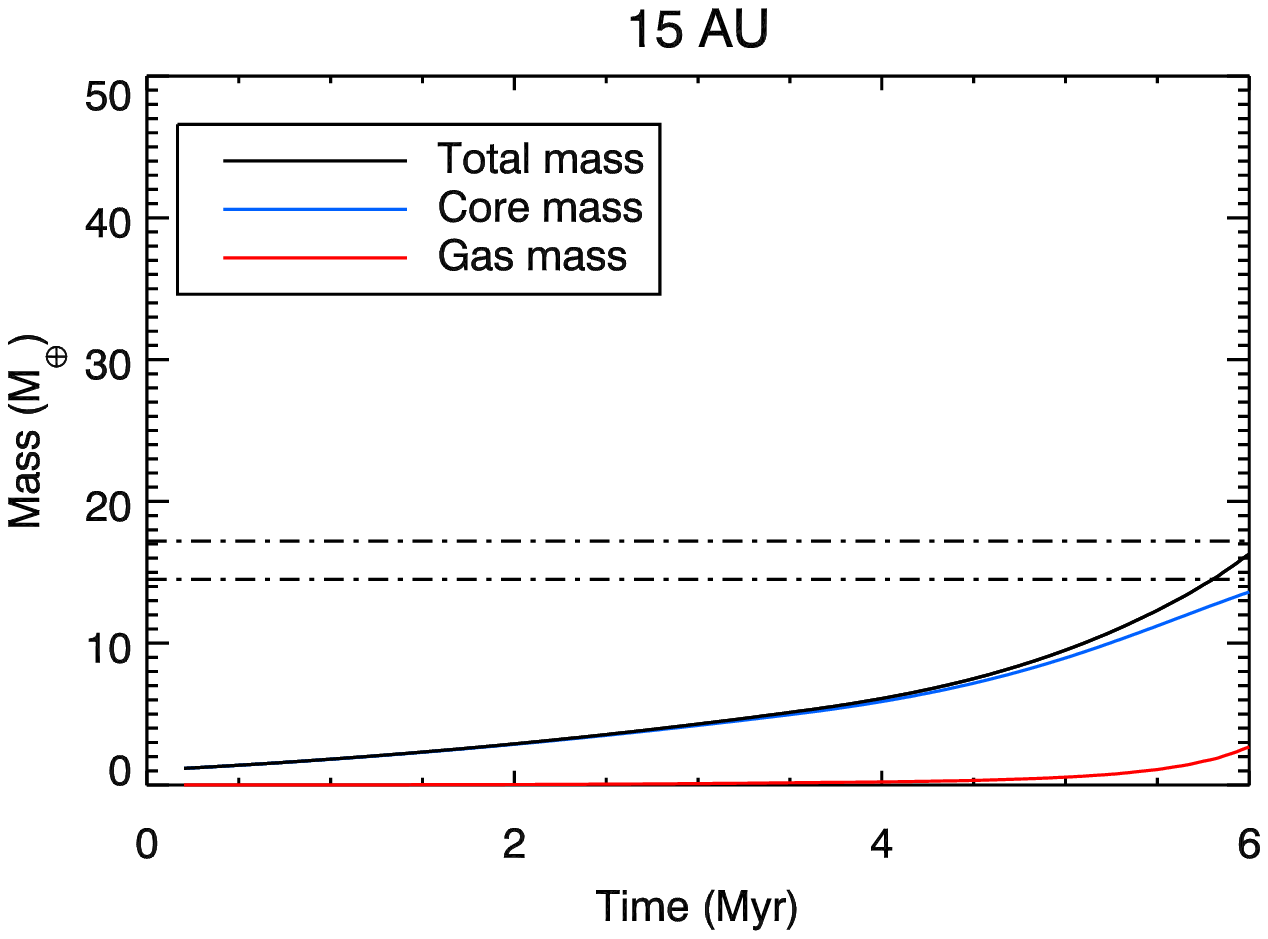}
\caption{
%{\bf Figure 2 Caption:}
Planetesimal ejection by the growing ice giant cores leads to
self-limiting accretion, which removes the possibility of runaway solid
growth and only slightly increases the accretion timescales for both
planets.  Left: The innermost ice giant, forming at 12~AU, attains its
current mass in 4.0~Myr, as compared to 3.8~Myr when planetesimal
ejection is not included in the simulation. Rapid gas accretion begins
as soon as Planet I's solid growth rate begins to level off, at the
inflection point of the $M_{\rm core}(t)$ curve. Right: At 15 AU, the
outer-forming ice giant reaches its present mass in 6~Myr, vs. 5.3~Myr
without planetesimal ejection. The dashed lines show the planets'
current masses. 
}
\label{ejectplot}
\end{figure}

\begin{figure}
\plottwo{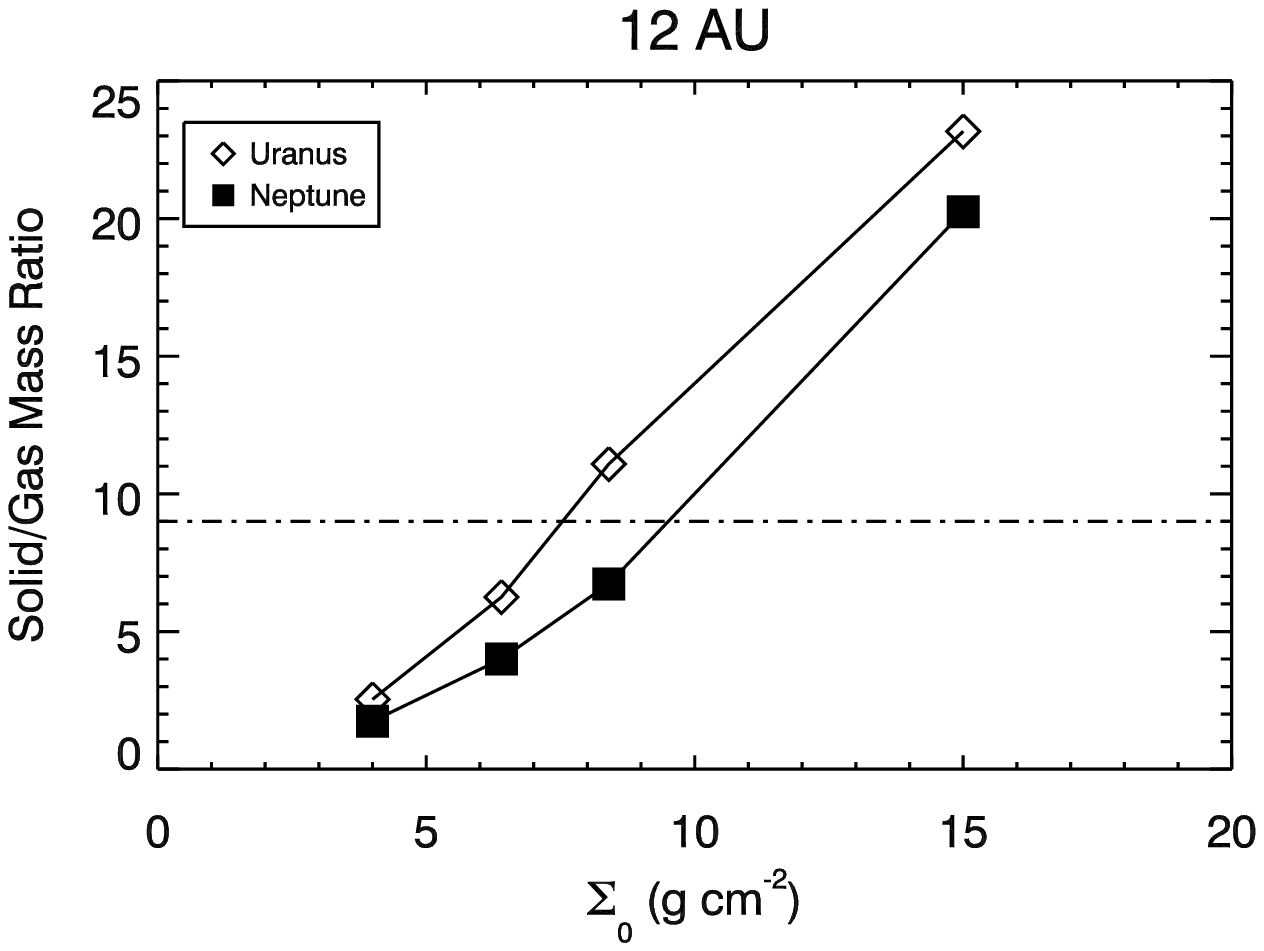}{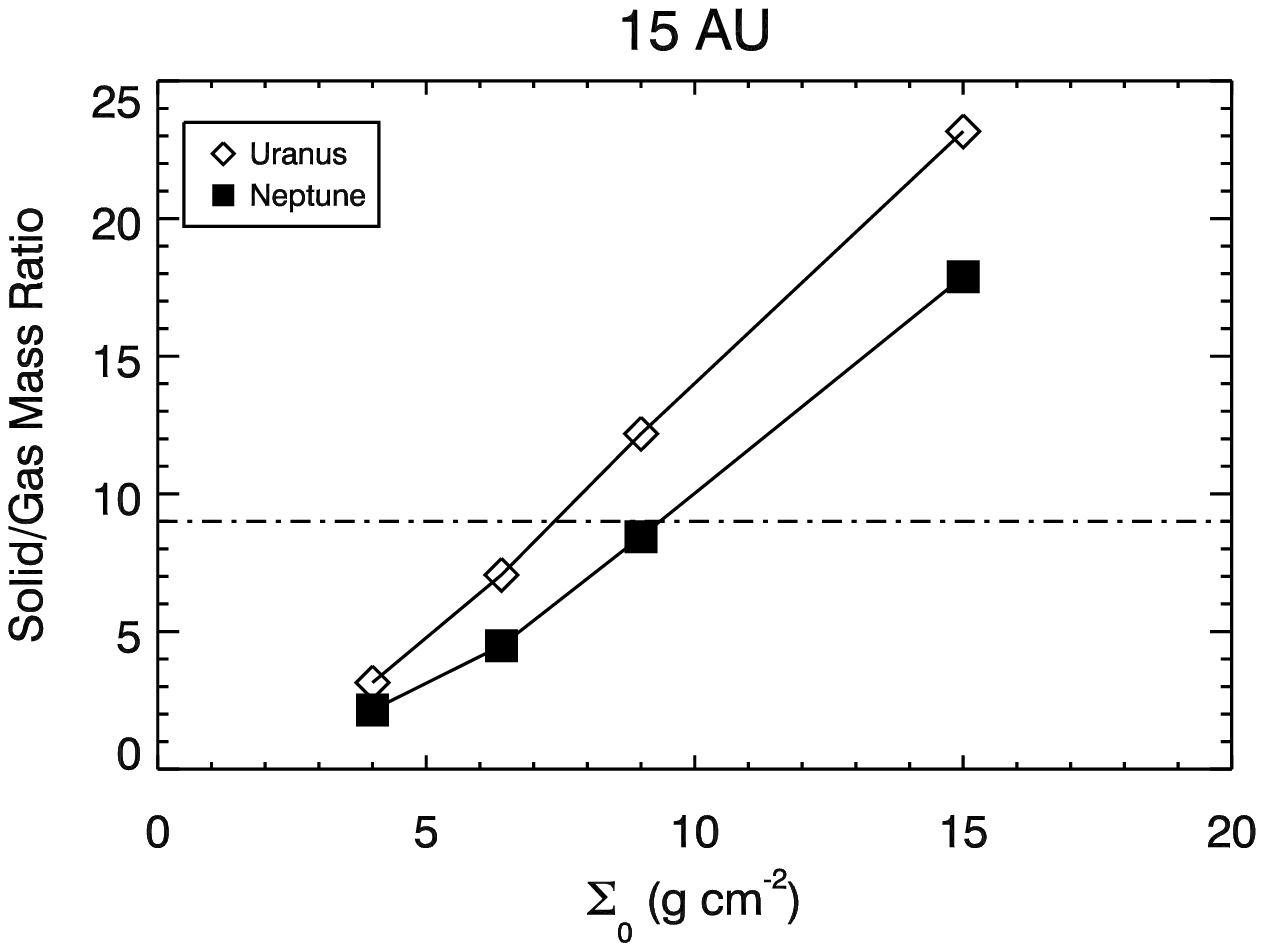}
\caption{
%{\bf Figure 3 Caption:}
Final solid/gas mass ratio in the planet is a strong function
of initial solid surface density, $\Sigma_0$, in the solar nebula. Both
Planet I and Planet O must have between 6~and~11~g~cm$^{-2}$ of
planetesimals in their feeding zones in order to reach the true ice
giant solid/gas ratio of $\sim 9$ (dash-dot line). The final solid/gas
ratio depends on whether the planet being formed is Uranus ($14.5
M_{\oplus}$) or Neptune ($17.2 M_{\oplus}$). High initial solid surface
densities lead to high solid accretion rates throughout formation, which
stabilize the protoplanet atmosphere against rapid contraction and lead
to the bona fide ice giant composition.
}
\label{solidgas}
\end{figure}

\end{document}